\pdfoutput=1
\documentclass[authorversion,nonacm]{acmart}
\AtBeginDocument{%
  \providecommand\BibTeX{{%
    \normalfont B\kern-0.5em{\scshape i\kern-0.25em b}\kern-0.8em\TeX}}}

\usepackage{url}
\usepackage{enumitem}
\usepackage{pgfplots}
\usepackage{multirow}
\usepackage{graphicx}
\usepackage{adjustbox}
\usepackage{makecell}

\pgfplotsset{compat=1.18}

\begin{document}

\title{Knowledge Sharing in Manufacturing using Large Language Models: User Evaluation and Model Benchmarking}


\author{Samuel Kernan Freire}
\orcid{0001-8684-0585}
\affiliation{%
  \institution{Delft University of Technology}
  \streetaddress{Landbergstraat 15}
  \city{Delft}
  \postcode{2628 CE}
  \country{The Netherlands}}
\email{s.kernanfreire@tudelft.nl}

\author{Chaofan Wang}
\orcid{0001-8213-6582}
\affiliation{%
  \institution{Delft University of Technology}
  \streetaddress{Landbergstraat 15}
  \city{Delft}
  \postcode{2628 CE}
  \country{The Netherlands}}
\email{c.wang-16@tudelft.nl}

\author{Mina Foosherian}
\orcid{0000-0002-2399-6213}
\affiliation{%
  \institution{BIBA - Bremer Institut für Produktion und Logistik GmbH}
  \streetaddress{Hochschulring 20}
  \city{Bremen}
  \country{Germany}
  \postcode{28359}}
\email{fos@biba.uni-bremen.de}

\author{Stefan Wellsandt}
\orcid{0000-0002-0797-0718}
\affiliation{%
  \institution{BIBA - Bremer Institut für Produktion und Logistik GmbH}
  \streetaddress{Hochschulring 20}
  \city{Bremen}
  \country{Germany}
  \postcode{28359}}
\email{wel@biba.uni-bremen.de}

\author{Santiago Ruiz-Arenas}
\orcid{0002-4018-7370}
\affiliation{%
  \institution{Universidad EAFIT}
  \streetaddress{Carrera 49 N° 7 Sur-50}
  \city{Medellin}
  \postcode{}
  \country{Colombia}}
\email{sruizare@eafit.edu.co}

\author{Evangelos Niforatos}
\orcid{0000-0002-0484-4214}
\affiliation{%
  \institution{Delft University of Technology}
  \streetaddress{Landbergstraat 15}
  \city{Delft}
  \country{Netherlands}}
\email{e.niforatos@tudelft.nl}

\renewcommand{\shortauthors}{Kernan Freire, et al.}

\begin{abstract}
Recent advances in natural language processing enable more intelligent ways to support knowledge sharing in factories. In manufacturing, operating production lines has become increasingly knowledge-intensive, putting strain on a factory's capacity to train and support new operators. This paper introduces a Large Language Model (LLM)-based system designed to retrieve information from the extensive knowledge contained in factory documentation and knowledge shared by expert operators. The system aims to efficiently answer queries from operators and facilitate the sharing of new knowledge. We conducted a user study at a factory to assess its potential impact and adoption, eliciting several perceived benefits, namely, enabling quicker information retrieval and more efficient resolution of issues. However, the study also highlighted a preference for learning from a human expert when such an option is available. Furthermore, we benchmarked several commercial and open-sourced LLMs for this system. The current state-of-the-art model, GPT-4, consistently outperformed its counterparts, with open-source models trailing closely, presenting an attractive option given their data privacy and customization benefits. In summary, this work offers preliminary insights and a system design for factories considering using LLM tools for knowledge management.

\end{abstract}

\maketitle

\section{Introduction}
Human-centric manufacturing seeks to support human operators with technology, aiming to enhance creativity, human well-being, problem-solving abilities, and overall productivity within factories~\cite{Alves2023,mayNewHumancentricFactory2015,FANTINI2020105058}. However, a significant challenge persists in effectively managing and utilizing the vast knowledge generated within these manufacturing environments, such as issue reports and machine documentation~\cite{ManufacturingKnowledgeRepository2014}. This knowledge is crucial for optimizing operations, yet it remains largely untapped due to the difficulties in processing and interpreting the disconnected, sometimes unstructured, technical information it contains~\cite{leoniMediatingRoleKnowledge2022,edwards2008clustering}.

Traditionally, leveraging this knowledge has been cumbersome, with operators choosing to use personal smartphones over official procedures~\cite{richterKnowledgeManagementDark2019} and AI unable to handle the complexity of the data~\cite{edwards2008clustering}. However, recent Large Language Models (LLMs) like GPT-4 show promise in addressing these challenges. LLMs can effectively interpret, summarize, and retrieve information from vast text-based datasets~\cite{lewis2020retrieval} while concurrently aiding the capture of new knowledge~\cite{kernan2023tacit}. These capabilities could significantly support operators in knowledge-intensive tasks, making it easier to access relevant information, share new knowledge, and make informed decisions rapidly.

While LLMs offer promising capabilities, their application in manufacturing is not straightforward. The specific, dynamic knowledge required in this domain poses unique challenges~\cite{Feng2017TowardKM}. For instance, a foundational LLM may have limited utility in a factory setting without significant customization, such as fine-tuning or incorporating specific context information into its prompts~\cite{Wang2023EmpowerLL}. Additionally, the practical and socio-technical risks and challenges of deploying LLMs in such environments remain largely unexplored --- factors key to human-centered AI~\cite{shneiderman2022human}. Concerns include the accuracy of the information provided, the potential for "hallucinated" answers~\cite{zuccon2023chatgpthaluc}, and the need for systems that can adapt to the highly specialized and evolving knowledge base of a specific manufacturing setting~\cite{Feng2017TowardKM}.

In response to these challenges, we developed an LLM-powered tool to leverage factory documents and issue analysis reports to answer operators' queries. Furthermore, the tool facilitates the analysis and reporting of new issues. This tool demonstrates the feasibility of using LLMs to enhance knowledge management in manufacturing settings. To understand its effectiveness and potential, we conducted a user study in a factory environment, evaluating the system's usability, user perceptions, adoption, and impact on factory operations.

Our approach also addresses the lack of specific benchmarks for evaluating LLMs in manufacturing. We benchmarked several LLMs, including both closed and open-source models, recognizing that the standard benchmarks\footnote{\url{https://huggingface.co/spaces/HuggingFaceH4/open_llm_leaderboard}---last accessed \today} primarily focus on general knowledge and reasoning. As such, they may not adequately reflect the challenges of understanding manufacturing-specific terminology and concepts. This benchmarking focused on their ability to utilize factory-specific documents and unstructured issue reports to provide factual and complete answers to operators' queries.


\section{Background}
In this section, we address the topic of industry 5.0, LLM-powered tools for knowledge management, benchmarking LLMs, and the research questions informing this work.
\subsection{Human-centered Manufacturing}
Industry 5.0, the latest phase of industrial development, places human beings at the forefront of manufacturing processes, emphasizing their skills, creativity, and problem-solving abilities~\cite{xu2021industry,maddikunta2022industry,Alves2023}. 
Human-centered manufacturing in Industry 5.0 focuses on providing a work environment that nurtures individuals' creativity and problem-solving capabilities~\cite{Madd2022}. It encourages workers to think critically, innovate, and continuously learn. With machines handling repetitive and mundane tasks, human workers can dedicate their time and energy to more complex and intellectually stimulating activities. This shift could enhance job satisfaction and promote personal and professional growth, as workers could acquire new skills and engage in higher-level decision-making~\cite{Alves2023,XuLu2021}. Emphasis on human-machine collaboration and the continuous emergence and refinement of technology increases the need for adequate human-computer interaction~\cite{Bru2023}. One of the approaches to address this topic is using conversational AI to assist humans in manufacturing~\cite{Well2021}.

\subsection{LLM-powered Knowledge Management Tools}
Training Large Language Models (LLMs) on numerous, diverse texts results in the embedding of extensive knowledge~\cite{zhao2023survey}. LLMs can also adeptly interpret complex information~\cite{jawahar-etal-2019-bert}, general reasoning~\cite{wei2022emergent}, and aiding knowledge-intensive decision-making. Consequently, researchers have been exploring applying LLM-powered tools in domain-specific tasks~\cite{Xie2023DARWINSD,Wen2023ChatHomeDA,Zhang2023BridgingTI}. 

Despite their potential benefits, the responses generated by LLMs may have two potential issues: (1) outdated information originating from the model's training date, and (2) inaccuracies in factual representation, known as ``hallucinations''~\cite{bang2023multitask, zhao2023survey}. To address these challenges and leverage the capabilities of LLMs in domain-specific knowledge-intensive tasks, several techniques can be used, such as chain-of-thought~\cite{wei2022chain}, few-shot prompting~\cite{brown2020, gao-etal-2021-making}, and retrieval augmented generation~\cite{lewis2020retrieval}.


Using few-shot prompting to retrieve information across diverse topics,~\citet{semnani2023wikichat} introduced an open-domain LLM-powered chatbot called WikiChat. WikiChat utilizes a 7-stage pipeline of few-shot prompted LLM that suggests facts verified against Wikipedia, retrieves additional up-to-date information, and generates coherent responses. They used a hybrid human-and-LLM method to evaluate the chatbot on different topics for factuality, alignment with real-worth truths and verifiable facts, and conversationality. This compound metric scores how informational, natural, non-repetitive, and temporally correct the response is. Their solution significantly outperforms GPT-3.5 in factuality, with an average improvement of 24.4\% while staying on par in conversationality. Others have explored the capabilities of LLMs in domain-specific tasks such as extracting structured data from unstructured healthcare texts~\cite{tang2023does}, providing medical advice~\cite{nov2023putting}, simplifying radiology reports~\cite{jeblick2022chatgpt}, Legal Judgement Prediction from multilingual legal documents~\cite{trautmann2022legal}, and scientific writing~\cite{Alkaissi.2023}.

Several manufacturers are cautiously adopting LLMs, while seeking solutions to mitigate their associated risks. For example,~\citet{mercedes2023} used AI with ChatGPT integrated through Azure OpenAI Service to enhance quality management and process optimization in vehicle production. This AI-driven approach simplifies complex evaluations for quality engineers through dialogue-based queries. \citet{xia2023autonomous} demonstrated how using in-context learning and injecting task-specific knowledge into an LLM can streamline intelligent planning and control of production processes.~\citet{kernan2023harnessing} built a proof of concept for bridging knowledge gaps among workers by utilizing domain-specific texts and knowledge graphs. \citet{WANG20237} conducted a systematic test of ChatGPT's responses to 100 questions from course materials and industrial documents. They used a zero-shot method and examined the responses' correctness, relevance, clarity, and comparability. Their results suggested areas for improvement, including low scores when responding to critical analysis questions, occasional non-factual or out-of-manufacturing scope responses, and dependency on query quality. Although~\cite{WANG20237} provides a comprehensive review of ChatGPT's abilities to answer questions related to manufacturing; it did not include the injection of task-specific knowledge into the prompts.

To improve the performance of an LLM for domain-specific tasks, relevant context information can be automatically injected along with a question prompt. This technique, known as Retrieval Augmented Generation (RAG), involves searching a corpus for information relevant to the user's query and inserting it into a query template before sending it to the LLM~\cite{lewis2020retrieval}. Using RAG also enables further transparency and explainability of the LLM's response. Namely, users can check the referenced documents to verify the LLM's response. Factories will likely have a large corpus of knowledge available in natural language, such as standard work instructions or machine manuals. Furthermore, factory workers continually add to the pool of available knowledge through (issue) reports. Until recently, these reports were considered unusable by AI natural language processing due to quality issues such as poorly structured text, inconsistent terminology, or incompleteness~\cite{edwards2008clustering}. However, the leap in natural language understanding that LLMs, such as ChatGPT, have brought about can overcome these issues.

\subsection{Evaluating LLMs}
Large Language Model evaluation requires the definition of evaluation criteria, metrics, and datasets associated with the system's main tasks. There are two types of LLM evaluations: intrinsic and extrinsic evaluation. Intrinsic evaluation focuses on the internal properties of a Language Model~\cite{Wei2023}. It means the patterns and language structures learned during the pre-training phase. Extrinsic evaluation focuses on the model’s performance in downstream tasks, i.e., in the execution of specific tasks that make use of the linguistic knowledge gained upstream, like code completion~\cite{Xu2022}. Despite extrinsic evaluation being computationally expensive, only conducting intrinsic evaluation is not comprehensive, as it only tests the LLMs capability for memorization~\cite{jang2022}. Here, we focus on extrinsic evaluation as we are primarily interested in the performance of LLM-based tools for specific real-world tasks. 


Extrinsic evaluation implies assessing the systems’s performance in tasks such as question answering, translation, reading comprehension, and text classification, among others~\cite{Kwon2022}. Existing benchmarks such as LAMBADA, HellaSwag, TriviaQA, BLOOM, Galactica, ClariQ and MMLU, among others, are widely reported in the literature for comparing language models. Likewise, domain-specific Benchmarks for tasks such as medical~\cite{singhal2022large}, fairness evaluation~\cite{Zhang2023}, finance~\cite{Xie2023}, robot policies~\cite{Liang2022}, and 3D printing code generation~\cite{Badini2023} can also be found. Experts also evaluate the performance of large-language models (LLMs) in specific downstream tasks, such as using physicians to evaluate the output of medical specific LLMs~\cite{singhal2022large}.

LLM benchmarks range from specific downstream tasks to general language tasks. However, to our knowledge, LLMs have not been benchmarked for answering questions in the manufacturing domain based on context material, a technique known as Retrieval Augmented Generation~\cite{lewis2020retrieval}. Material such as machine documentation, standard work instructions, or issue reports will contain domain jargon and technical information that LLMs may struggle to process. Furthermore, the text in an issue report may pose additional challenges due to abbreviations, poor grammar, and formatting~\cite{edwards2008clustering}. Therefore, as part of this work, we benchmarked several LLMs on their ability to answer questions based on factory manuals and unstructured issue reports. Furthermore, we conducted a user study with factory operators and managers to assess the potential benefits, risks and challenges. The following research questions informed our study:

\begin{enumerate}
    \item \textit{What are the perceived benefits, challenges, and risks of using Large Language Models for information retrieval and knowledge sharing for factory operators?}\label{rq:1} 
    \item \textit{How do Large Language Models compare in performance when answering factory operators' queries based on factory documentation and unstructured issue reports?}\label{rq:2} We consider performance as the factuality, completeness, hallucinations, and conciseness of the generated response. 
\end{enumerate}

\section{System}
\begin{figure}[ht!]
\begin{center}
\includegraphics[width=16cm]{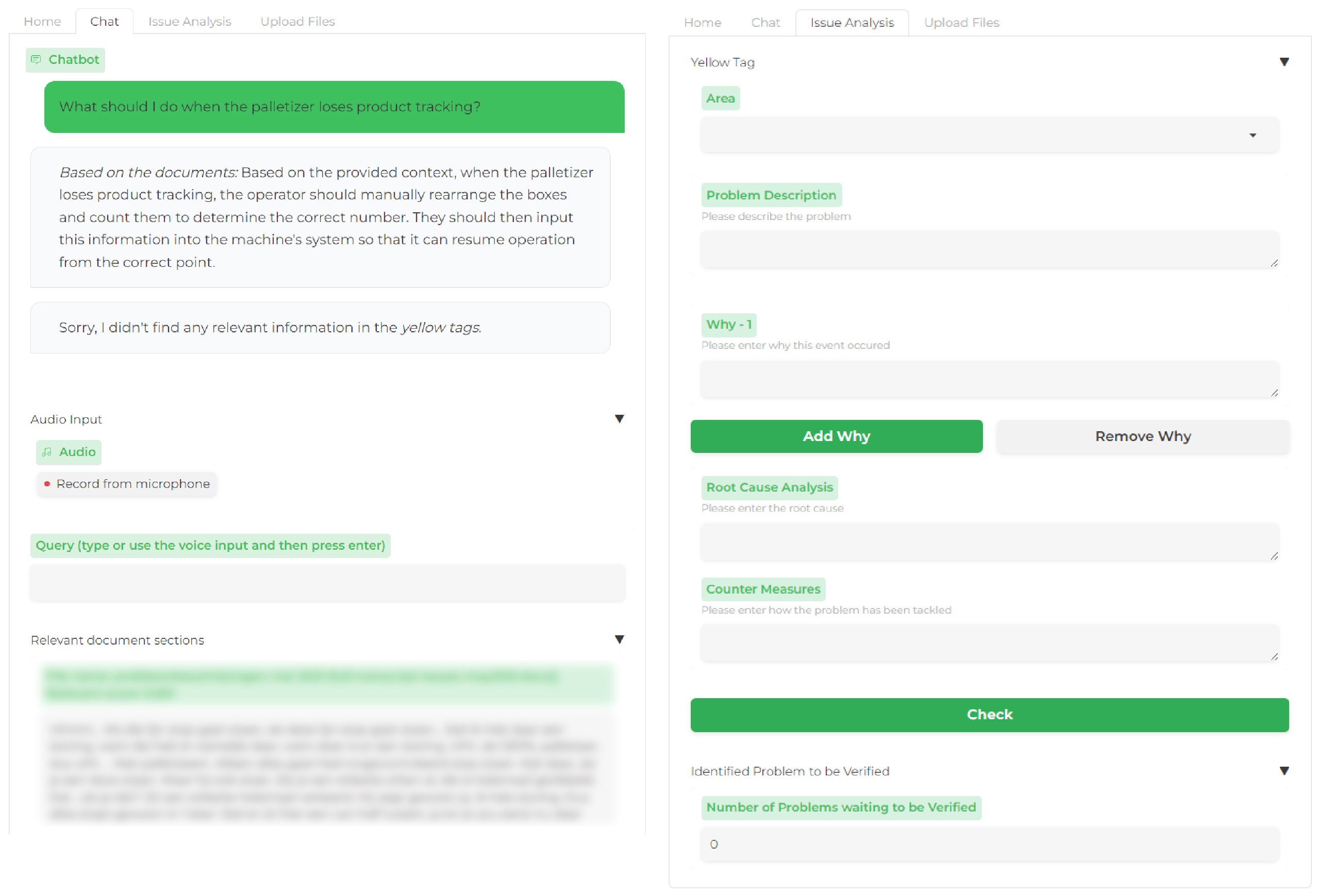}
\end{center}
\caption{The main screens for the tool's interface are the chat interface and issue analysis screen. The ``relevant document sections'' part is blurred for confidentiality as it shows the title of a company's document and its content.}\label{fig:interface}
\end{figure}

We built a fully functional system to assess the potential of using LLMs for information retrieval and knowledge sharing for factory operators. Benefiting from LLMs' in-context learning capabilities, we use this to supply an LLM with information in the form of factory manuals, and issue reports relevant to the user's question, a technique known as Retrieval Augmented Generation (RAG)~\cite{lewis2020retrieval}, see Figure~\ref{fig:RAG}. As noted by~\citet{wei2022emergent}, training LLMs using a prompt packed with query-related information can yield substantial performance enhancement. Users can ask questions in the chat box by typing or using voice input. The response is displayed at the top of the page, and the document chunks used for the answer can be checked at the bottom (see Figure~\ref{fig:interface}).

\begin{figure}[ht!]
\begin{center}
\includegraphics[width=10cm]{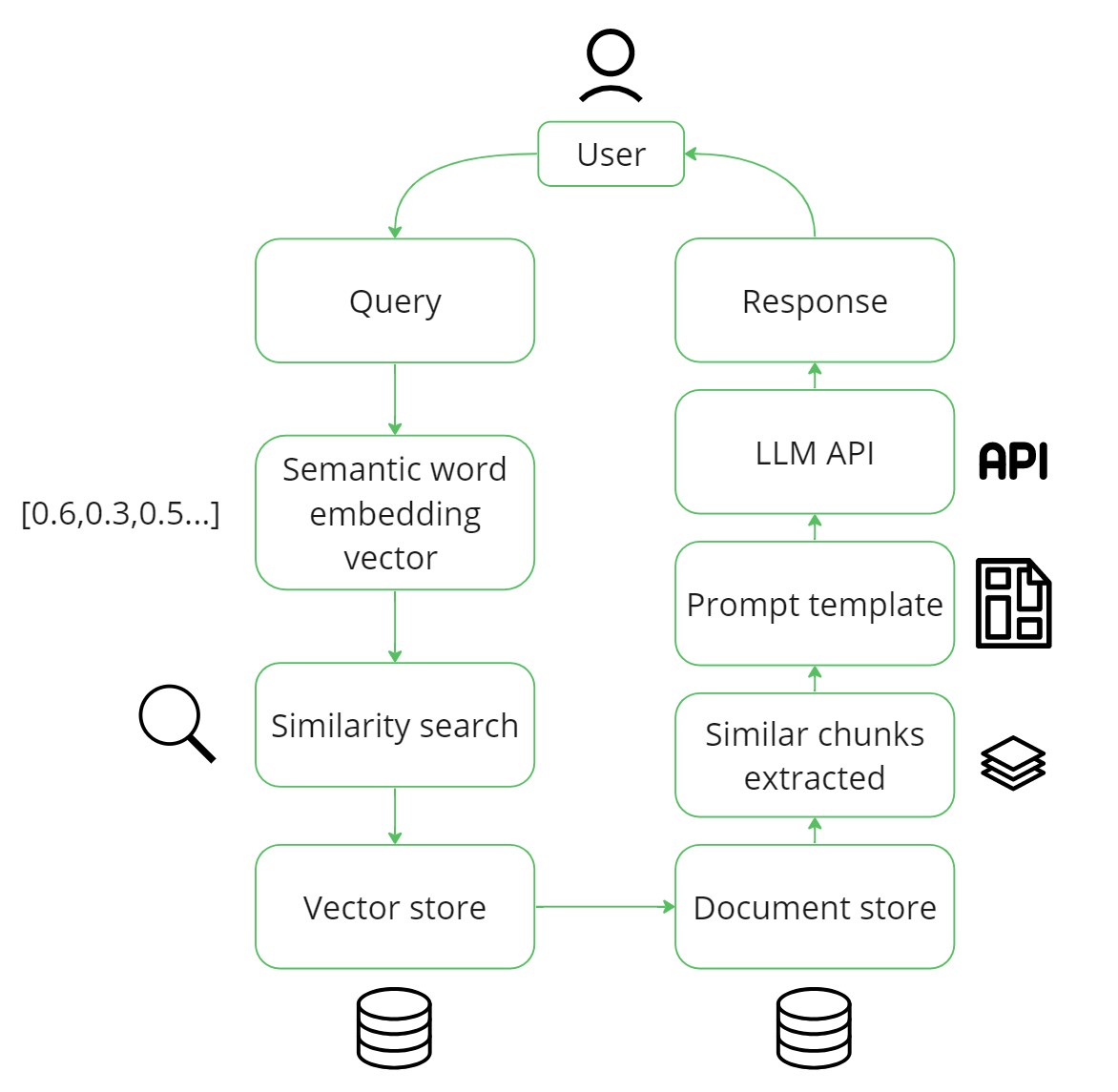}
\end{center}
\caption{The steps of Retrieval Augmented Generation (RAG) from user query to response}\label{fig:1}
\end{figure}\label{fig:RAG}

\subsection{Tool Dependencies}

The tool was constructed utilizing two innovative technologies - Gradio and LlamaIndex. Gradio, a tool developed by~\citet{Abid_Gradio_Hassle-free_sharing_2019}, serves as the backbone for both our front and back ends. Primarily used to simplify the development and distribution of machine learning applications, Gradio allows the quick creation of intuitive, user-friendly web interfaces for machine learning models.

Additionally, we use LlamaIndex, created by~\citet{Liu_LlamaIndex_2022}, for retrieving the context material in response to the user queries and handling the queries to the LLM. LlamaIndex, initially known as GPT Index, is a cutting-edge data framework designed for the efficient handling and accessibility of private or domain-specific data in LLMs applications.

Since the factory documents can be long, they may overflow the LLM's context window or result in unnecessary computational demand. To overcome this, we segment the materials into manageable chunks, each comprising approximately 400 tokens. This method effectively incorporates the materials into the LLM prompt without compromising the conversation flow. Following the segmentation, each document chunk is processed through LlamaIndex using the OpenAI Embedding API\footnote{\url{https://api.openai.com/v1/embeddings}---last accessed \today}. Utilizing the ``text-embedding-ada-002'' model, LlamaIndex transforms each chunk into a corresponding embedding vector. These resulting vectors are then securely stored, ready for future retrieval and use.

\subsection{Knowledge Base Construction}
Our experiment incorporates two distinct types of domain-specific data: factory manuals and shared knowledge from factory workers. Factory manuals outline information on machine operation, safety protocols, quality assurance, and more. These resources, provided by factory management teams, initialize the knowledge base for each specific factory. The materials come in various formats, including PDF, Word, and CSV files.

In addition to the factory manuals, we integrate issue analysis reports from factory workers. This information is gathered from the production line, utilizing the five-why process, an iterative root-cause analysis technique~\cite{serrat2017five} (right side of Figure~\ref{fig:interface}). The five-why technique probes into cause-and-effect relationships underlying specific problems by repeatedly asking "Why?" until the root cause is revealed, typically by the fifth query. This process enables us to gather real-world issues encountered on production lines, which may not be covered in the factory manuals. Upon entering all required information, including one or more ``whys'', the operator presses ``check'', triggering a prompt to the LLM that performs a logical check of the entered information and checks for inconsistencies with previously reported information. The operator can revise the entered information and submit it as is. Then, the submitted report will be added to a queue for expert operators to check before it is added to the knowledge base.

\subsection{Query Construction}
To retrieve the document data relevant to specific user queries, we employ the same embedding model, ``text-embedding-ada-002'' to generate vector representations of these queries. By leveraging the similarity calculation algorithm provided by LlamaIndex, we can identify and retrieve the top-K most similar segmented document snippets related to the user query. This allows us to construct pertinent LLM queries. Once the snippets are retrieved, they are synthesized into the following query template based on the templates used by LlamaIndex\footnote{\url{https://docs.llamaindex.ai}---last accessed \today}:
\begin{quote}
    You are an assistant that assists detergent production line operators with decision support and advice based on a knowledge base of standard operating procedures, single point lessons (SPL), etc. We have provided context information below from relevant documents and reports.

    \noindent\makebox[\linewidth]{\rule{0.8\textwidth}{1pt}}
    \centering{[Retrieved Document Snippets]}
    
    \noindent\makebox[\linewidth]{\rule{0.8\textwidth}{1pt}}
    \raggedright 

    Given this information, please answer the following question:
    \centering{[Query]}
    
    \raggedright
    If the provided context does not include relevant information to answer the question, please do not respond.
\end{quote}\label{prompt_template}

However, considering our data originates from two distinct sources – factory manuals and shared tactical knowledge – we have decided to segregate these into two separate LLM queries. This approach is designed to prevent potential user confusion from combining data from both sources into a single query.

\section{User Study in the Field}
We conducted a user study on the system to uncover perceived benefits, usability issues, risks, and barriers to adoption. The study comprised three tasks: to ask the system several questions as if they were operators, to fill in a ``yellow tag'' (issue analysis report) based on a recent issue and request a logical check, and finally, to upload new documents to the system. After each task, they were asked to provide feedback. Then, after completing all tasks, the participants were posed several open questions about the system's benefits, risks, and barriers to adoption. Finally, demographic information, such as age, gender, and role, was collected.

\subsection{Participants}
We recruited $N = 9$ participants from a detergent factory, of which $n = 4$ were managers (P1-4), and $n = 5$ were operators (P5-9). Of the nine participants, $n = 3$ were women, and $n = 6$ were men. Participant age was distributed over three brackets, namely $n = 2$ were 30--39, $n = 4$ were 40--49, and $n = 3$ were 50--59.

\subsection{Qualitative Analysis}
An inductive thematic analysis~\cite{guest2011applied} of the answers to the open questions resulted in six themes discussed below.

\begin{itemize}
    \item \textbf{Usability} The theme of usability underlines the system's ease of use and the need for clear instructions. Users mentioned the necessity for a ``user-friendly'' (P2) interface and highlighted the importance of having ``more instructions and more details need to be loaded'' (P1) to avoid confusion. This indicates a desire for intuitive navigation that could enable workers to use the system effectively without extensive training or referencing external help. The feedback suggests that the system already works well, as reflected in statements like ``Easy-to-use system'' (P3) and the system ``works well'' (P7).
    \item \textbf{Access to information} Users appreciated the ``ease of having information at hand'' (P1), facilitating immediate access to necessary documents. However, there is a clear call for improvements, such as the ability to ``Include the possibility of opening IO, SPL, etc. in .pdf format for consultation'' (P3). This theme is supported by requests for direct links to full documents, suggesting that while ``the list of relevant documents from which information is taken is excellent'' (P4), the ability to delve deeper into full documents would significantly enhance the user experience.
    \item \textbf{Efficiency} Users value the ``greater speed in carrying out some small tasks'' (P3). However, there are concerns about the system's efficiency when it does not have the answer, leading to ``wasting time looking for a solution to a problem in case it is not reported in the system's history'' (P3). Statements like ``quick in responses'' (P3) contrast with the need for questions to be ``too specific to have a reliable answer'' (P7), indicating tension between the desire for quick solutions and the system's limitations.
    \item \textbf{Adoption} Users highlight several factors affecting adopting the new system. It includes challenges such as ``awareness and training of operators [might hinder adoption]'' (P3) and the need for ``acceptance by all employees'' (P4), which indicates that the system's success is contingent on widespread user buy-in. The generational divide is also noted: ``That older operators use it [on what may hinder adoption]'' (P7) suggests that demographic factors may influence the acceptance of new technology.
    \item \textbf{Safety} Users express apprehension that ``if the responses are not adequate, you risk safety'' (P1), emphasizing the critical nature of reliable information in a factory setting. Moreover, the demand for updated and specific information underlines the importance of the system’s content being current and detailed to maintain operational safety standards, as stated by P9: ``If it is updated and specific, it can help me''.
    \item \textbf{Traditional versus Novel} There is a noticeable preference for established practices among some users. For instance, ``It's faster and easier to ask an expert colleague working near me rather than [the system]'' (P8) captures the reliance on human expertise over the assistant system. This tension is further demonstrated by the sentiment that ``Operators may benefit more from traditional information retrieval systems'' (P9), suggesting a level of skepticism or comfort with the status quo that the new system needs to overcome.
\end{itemize}

\section{LLM Benchmarking}
In our benchmarking experiment, we evaluated various commercial and open-source LLMs, including OpenAI's ChatGPT (GPT-3.5 and GPT-4 from July 20th 2023), Guanaco 65B and 35B variants~\cite{dettmers2023qlora} based on Meta's Llama (Large Language Model Meta AI)~\cite{touvron2023llama}, Mixtral 8x7b~\cite{jiang2024mixtral}, Llama 2~\cite{touvron2023Llama2}, and one of its derivatives, StableBeluga2~\cite{StableBelugaModels}. This selection represents the state-of-the-art closed-sourced models (e.g., GPT-4) and open-source models (e.g., Llama 2). We included the (outdated) Guanaco models to demonstrate the improvements in the open-source sphere over the past year.

We used a web UI for LLMs\footnote{\url{https://github.com/oobabooga/text-generation-webui/tree/main}---last accessed \today} to load and test the Mixtral 8x7B, Guanaco models, and the StableBeluga2. The models were loaded on a pair of Nvidia A6000s with NVlink and a total Video Random Access Memory (VRAM) capacity of 96 GB. The 65B model was run in 8-bit mode to fit in the available VRAM. We used the llama-precise parameter preset and fixed zero seed for reproducibility. Llama 2 was evaluated using the demo on huggingface\footnote{\url{https://huggingface.co/meta-llama/Llama-2-70b-chat-hf}---last accessed \today}.

To rigorously assess the models, we prepared 20 questions of varying complexity based on two types of context material: half from operating manuals and half from unstructured issue reports. The operating manuals included excerpts from actual machine manuals and standard operating procedures, while the informal issue reports were free-text descriptions of issues we had previously collected from operators. The model prompt was constructed using the above template (\ref{prompt_template}). Ultimately, the difficulty of a question is a combination of the question's complexity and the clarity of the source material. Simple questions include retrieving a single piece of information clearly stated in the context material, for example, "At what temperature is relubrication necessary for the OKS 4220 grease?". Conversely, difficult questions require more reasoning or comprise multiple parts, for example, "What should I do if the central turntable is overloaded?" which has a nuanced answer dependent on several factors not clearly articulated in the context material. 

In addition to measuring response length in words, every response is manually scored on factuality, completeness, and hallucinations as defined below:

\begin{itemize}
    \item \textbf{Factuality}: Responses align with the facts in the context material.
    \item \textbf{Completeness}: Responses contain all the information relevant to the question in the context material.
    \item \textbf{Hallucinations}: Response appears grammatically and semantically coherent but is not based on the context material.
\end{itemize}

The following scoring protocol is applied: one is awarded for a completely factual, complete, or hallucinated response. In contrast, a score of 0.5 is awarded for a slightly nonfactual, incomplete, or hallucinated response (e.g., the response includes four out of the five correct steps). Otherwise, a score of zero is awarded. Therefore, wrong answers are penalized heavily. If the model responds by saying it cannot answer the question and does not make any attempt to do so, it is scored zero for factuality and completeness, but no score is given for hallucination. As such, the final score for hallucination is calculated as follows: $\text{corrected score} = \frac{\text{score}}{20 - \text{number of unanswered questions}} \times 100$

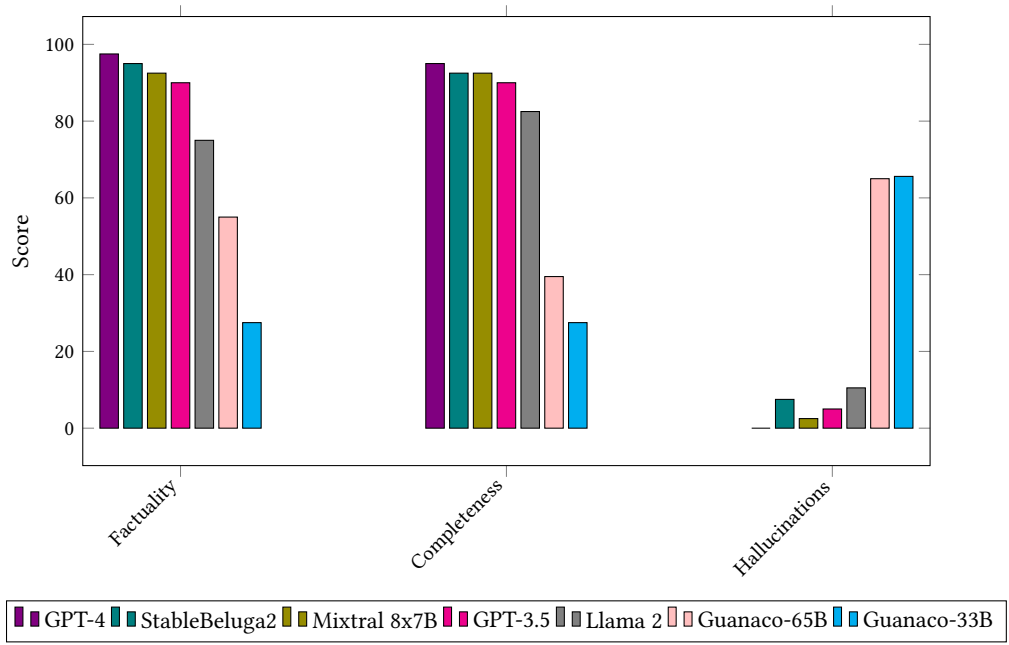
\begin{figure}[ht]
\centering
\begin{tikzpicture}
\begin{axis}[
    width=0.85\linewidth,
    height=0.5\linewidth,
    ybar,
    bar width=7pt,
    enlarge x limits=0.15,
    legend style={at={(0.5,-0.3)}, anchor=north, legend columns=-1}, 
    ylabel={Score},
    symbolic x coords={Factuality, Completeness, Hallucinations},
    xtick=data,
    nodes near coords align={vertical},
    x tick label style={rotate=45, anchor=east, font=\small},
    y tick label style={font=\small},
]
\addplot[fill=violet] coordinates {(Factuality,97.5) (Completeness,95) (Hallucinations,0)};
\addplot[fill=teal] coordinates {(Factuality,95) (Completeness,92.5) (Hallucinations,7.5)};
\addplot[fill=olive] coordinates {(Factuality,92.5) (Completeness,92.5) (Hallucinations,2.5)};
\addplot[fill=magenta] coordinates {(Factuality,90) (Completeness,90) (Hallucinations,5)};
\addplot[fill=gray] coordinates {(Factuality,75) (Completeness,82.5) (Hallucinations,10.5)};
\addplot[fill=pink] coordinates {(Factuality,55) (Completeness,39.5) (Hallucinations,65)};
\addplot[fill=cyan] coordinates {(Factuality,27.5) (Completeness,27.5) (Hallucinations,65.6)};
\legend{GPT-4, StableBeluga2, Mixtral 8x7B, GPT-3.5, Llama 2, Guanaco-65B, Guanaco-33B}
\end{axis}
\end{tikzpicture}
\caption{Benchmark of seven LLMs for generating answers based on factory materials.}
\label{fig:language_models}
\end{figure}

As shown in Figure \ref{fig:language_models} and Table \ref{table:benchmarking}, GPT-4 outperforms other models regarding factuality, completeness, and lack of hallucinations but is closely followed by StableBeluga2 and GPT-3.5. The Guanaco models, based on Llama 1, perform significantly worse. The conciseness of the responses showed a similar pattern, except that StableBeluga2 produced the shortest answers (58 words), followed closely by Mixtral 8x7B (66 words) and GPT-4 (69 words).

\begin{table}[ht]
\centering
\caption{Model Benchmarking Scores (out of 100) and Average Response Length}
\label{table:benchmarking}
\vspace{1em} 
\begin{tabular}{lcccc}
\textbf{Model} & \textbf{Factuality} & \textbf{Completeness} & \textbf{Hallucinations} & \textbf{Words} \\ 
GPT-4 & 97.5 & 95 & 0 & 69 \\ 
StableBeluga2 & 95 & 92.5 & 7.5 & 58 \\
Mixtral 8x7B & 92.5 & 92.5 & 2.5 & 66 \\
GPT-3.5 & 90 & 90 & 5 & 89 \\
Llama 2 & 77.5 & 82.5 & 13 & 128 \\ 
Guanaco 65B & 55 & 39.5 & 65 & 131 \\ 
Guanaco 33b & 27.5 & 27.5 & 65.6 & 190 \\ 
\end{tabular}
\end{table}

\section{Discussion}
\subsection{GPT-4 is the best, but open-source models follow closely}
GPT-4 performs best across all measures but is closely followed by StableBeluga2, Mixtral 8x7B, and GPT-3.5. Compared to GPT-4, the cost per input token for GPT-3.5 is significantly lower\footnote{\url{https://openai.com/pricing\#language-models}---last accessed \today}. However, the higher costs of GPT-4 are partially counteracted by its concise yet complete responses. If longer, more detailed responses were desired (e.g., for training purposes), the prompt could be adjusted. We observed that the less powerful models, such as GPT-3.5 and Llama 2, tended to be wordier and include additional details that were not directly requested. In contrast, GPT-4, StableBeluga2, and Mixtral 8x7B generated more concise responses.

The latest generation of open-source models, such as Mixtral 8x7B and Llama 2 variants, such as StableBeluga2, demonstrates a clear jump forward relative to their predecessors based on Llama-1, which were more prone to hallucinations and exhibited poorer reasoning abilities over the context material. While open-source models like StableBeluga2 and Mixtral 8x7B do not score as high as GPT-4, they ensure better data security, privacy, and customization if hosted locally. This can be a crucial consideration for companies with sensitive data or unique needs.

\subsection{The tool is beneficial but inferior to human experts}
Users appreciate the system's functionality and see it as a tool for modernizing factory operations and speeding up operations. They are keen on improvements to be made for better user experience and utility, especially in the areas of content, feature enhancements, and user training. However, they express concerns about potential safety risks and the efficacy of information retrieval compared to consulting expert personnel. While these concerns are understandable, the tool was not designed to replace human-human interactions; instead, it can be used when no human experts are present or when they do not know or remember how to solve a specific issue. This would come into play during the night shift at the factory where we conducted the user study as a single operator operates a production line, leaving limited options for eliciting help from others.

\subsection{Limitations and future work}
We used the same prompt for all LLMs; however, it is possible that some of the LLMs would perform better with a prompt template developed explicitly for it. However, we matched the LLMs' hyperparameters (e.g., temperature) as closely as possible across all the tested models, except for Llama 2, as we did not have access to the presets as we did not host it locally. Our model benchmarking procedure involved 20 questions, and a singular coder assessed the responses. This introduces the potential for bias, and the limited number of questions may not cover the full spectrum of complexities in real-world scenarios. However, we varied query complexity and source material types to (partially) mitigate these shortcomings.

The study's design did not include a real-world evaluation involving end users operating the production line, as this was considered too risky for our industry partner. Such an environment might present unique challenges and considerations not addressed in this research, such as time pressure. However, by involving operators and managers and instructing them to pose several questions based on their actual work experience, we could still evaluate the system and collect valid feedback.

These limitations suggest directions for future research, for example, longitudinal studies where operators use the tool during production line operations and more comprehensive prompt and model customization. Longitudinal studies will be key to understanding the real-world impact on production performance, operator well-being, and cognitive abilities.

\section{Conclusion}
The results demonstrated GPT-4's superior performance over other models regarding factuality, completeness, and minimal hallucinations. Interestingly, open-source models like StableBeluga2 and Mixtral 8x7B followed close behind. The user study highlighted the system's user-friendliness, speed, and logical functionality. However, improvements in the user interface and content specificity were suggested, along with potential new features. Benefits included modernizing factory operations and speeding up specific tasks, though concerns about safety, efficiency, and inferiority to asking human experts were raised.




\begin{acks}
This work was supported by the European Union’s Horizon 2020 research and innovation program via the project COALA “COgnitive Assisted agile manufacturing for a LAbor force supported by trustworthy Artificial Intelligence” (Grant agreement 957296).    
\end{acks}

\bibliographystyle{ACM-Reference-Format}
\bibliography{references}

\appendix

\end{document}